\pdfoutput=1

\documentclass[notitlepage,twocolumn,superscriptaddress, aps, prl, floatfix]{revtex4-1}

\usepackage{lmodern}
\usepackage{textcomp}
\usepackage{graphicx}
\usepackage{amsmath}
\usepackage{mathtools}
\usepackage{graphicx}
\usepackage{bm}
\usepackage{array}
\usepackage{dcolumn}
\usepackage{hepparticles}
\usepackage{heppennames}
\usepackage{hepnicenames}
\usepackage{lineno}
\usepackage[english]{babel}
\usepackage[autostyle]{csquotes}
\usepackage{physics}
\usepackage{qcircuit}
\usepackage{epstopdf}
\usepackage[colorlinks,
            linkcolor=blue,
            anchorcolor=blue,
            citecolor=green
            ]{hyperref}
\usepackage{subfigure}
\usepackage{nicefrac}

\begin{document}

\newcommand{\Yb}{$^{171}{\rm{Yb}}^{+} $}

\title{Two-qubit entangling gates within arbitrarily long chains of trapped ions}

\author{K. A. Landsman}\email{kevinlandsman@gmail.com}
\affiliation{These authors contributed equally}
\affiliation{Joint Quantum Institute, Department of Physics and Joint Center for Quantum Information and Computer Science, University of Maryland, College Park, MD 20742}
\author{Y. Wu}
\affiliation{These authors contributed equally}
\affiliation{Department of Physics, University of Michigan, Ann Arbor, MI 48109}
\author{P. H. Leung}
\affiliation{Department of Physics, Duke University, Durham, North Carolina 27708, USA}
\author{D. Zhu}
\affiliation{Joint Quantum Institute, Department of Physics and Joint Center for Quantum Information and Computer Science, University of Maryland, College Park, MD 20742}
\author{N. M. Linke} 
\affiliation{Joint Quantum Institute, Department of Physics and Joint Center for Quantum Information and Computer Science, University of Maryland, College Park, MD 20742}
\author{K. R. Brown}
\affiliation{Department of Physics, Duke University, Durham, North Carolina 27708, USA}
\affiliation{Department of Electrical and Computer Engineering, Duke University, Durham, North Carolina 27708, USA}
\author{L. Duan}
\affiliation{Center for Quantum Information, IIIS, Tsinghua University, Beijing 100084, China}
\affiliation{Department of Physics, University of Michigan, Ann Arbor, MI 48109}
\author{C. Monroe}
\affiliation{Joint Quantum Institute, Department of Physics and Joint Center for Quantum Information and Computer Science, University of Maryland, College Park, MD 20742}
\affiliation{IonQ, Inc., College Park, MD 20740}

\begin{abstract}
Ion trap systems are a leading platform for large scale quantum computers. Trapped ion qubit crystals are fully-connected and reconfigurable, owing to their long range Coulomb interaction that can be modulated with external optical forces. However, the spectral crowding of collective motional modes could pose a challenge to the control of such interactions for large numbers of qubits. Here, we show that high-fidelity quantum gate operations are still possible with very large trapped ion crystals, simplifying the scaling of ion trap quantum computers. To this end, we present analytical work that determines how parallel entangling gates produce a crosstalk error that falls off as the inverse cube of the distance between the pairs.
We also show experimental work demonstrating entangling gates on a fully-connected chain of seventeen \Yb ions with fidelities as high as $97(1)\%$.
\end{abstract}
\maketitle

\section{Introduction}
The central challenge in scaling a quantum computer is to increase the entangling quantum gate performance while more qubits are added to the system. Trapped ion qubits are well known leaders in both coherence properties \cite{wang_single-qubit_2017} and entangling gate fidelity \cite{gaebler_high-fidelity_2016,harty_high-fidelity_2016}. This allows ion trap systems to scale because their atomic clock qubits are almost perfectly replicable and have negligible idle errors \cite{monroe_scaling_2013}. Trapped ion qubits also have long-range interaction graphs \cite{monz_14-qubit_2011,debnath_demonstration_2016}, provided by optical forces that modulate the Coulomb-coupled motion of a crystal of ions \cite{cirac_quantum_1995,molmer_multiparticle_1999,solano_deterministic_1999,milburn_ion_2000}. 
Owing to the added complexity of the motion of large chains of trapped ion qubits, it might be expected that the speed or control of gates might be compromised. In this letter, we show there is no fundamental difficulty in extending high-fidelity entangling gates to arbitrarily long chains of ions, as the interactions can take on a local character.

Quantum entangling gates between trapped ion qubits in a single crystal or chain are mediated by the Coulomb-collective phonon modes of motion through qubit state-dependent forces. Each phonon mode is densely connected to all qubits via bosonic quasiparticles \footnote{except the middle ion in a chain with an odd number of identical ions}, allowing any qubit to be entangled with any other qubit in the crystal. However, this requires that each phonon mode be disentangled with the qubits after the gate operation. One way to accomplish this is to resolve just a single mode of motion that mediates the interaction \cite{cirac_quantum_1995}, but this generally slows the gate speed for larger ion crystals due to spectral crowding of the normal modes. Alternatively, many modes can be used to maintain gate speed, requiring that the optical force be modulated in a particular way to disentangle each of the modes after the gate operation \cite{zhu_arbitrary-speed_2006, lin_large-scale_2009, choi_optimal_2014, green_phase-modulated_2015, debnath_demonstration_2016, leung_robust_2018}.

In the regime where many modes are excited during a gate, the force envelope is pre-calculated by designing and optimizing a pulse shape that is constrained to produce the desired entanglement and decouple the qubits from the motional modes. This approach increases the classical complexity of the optical force pattern required to perform the gate in the presence of $n$ modes, ranging between $\mathcal{O}(n)$ to $\mathcal{O}(n^2)$ \cite{choi_optimal_2014, green_phase-modulated_2015}, although there are also methods that scale independently of system size \cite{leung_robust_2018}. In addition, as ions are added to the crystal, the transverse phonon modes become more tightly packed around the highest frequency mode. This increases the sensitivity of the gate fidelity to noise or drifts in the phonon mode frequencies. Here we show that by considering only a local set of ions, high-fidelity parallel gates can be performed even on an infinitely long chain. We also present experimental results on a 17-ion chain with small inter-ion distances where frequency-crowding problems are circumvented.

We create an Ising-type interaction \cite{molmer_multiparticle_1999,solano_deterministic_1999,milburn_ion_2000} by off-resonantly
driving phonon modes of the ionic crystal near sidebands of transverse motion. The qubit-phonon Hamiltonian in general takes the form \cite{lee_phase_2005}
\begin{equation}
H=\sum_{j,m} \eta_{j,m} f_j(t) \left( a_m^\dag e^{i\omega_m t} + a_m e^{-i\omega_m t} \right) \sigma_j^x
\end{equation}
for ion $j$ and collective oscillation mode $m$ along the laser's 
wavevector $\mathbf{k}$ (for two-photon stimulated Raman forces with wavevectors $\mathbf{k}_1$ and $\mathbf{k}_2$, $\mathbf{k}= \mathbf{k}_2 - \mathbf{k}_1$).  The Lamb-Dicke parameter of ion $j$ with mode $m$ is $\eta_{j,m} = b_{j,m}\sqrt{\hbar k^2 /(2m\omega_m)}$ where $b_{j,m}$ is the normal mode participation eigenvector and $k=|\bf{k}|$. The time-dependent laser forces on ion $j$ is characterized by the Rabi frequency $f_j(t)$, and $\sigma^x$ is the Pauli spin-flip operator in the x-basis. The above Hamiltonian presumes that the ions are confined within the Lamb-Dicke limit, but the effects of higher order terms in $\eta_{j,m}$ can be systematically bounded \cite{wu_noise_2018}.


The evolution of the above Hamiltonian after time $\tau$ follows the unitary operator
\begin{equation}
U = \exp\left( \sum_j \phi_j \sigma_j^x + i\sum_{i<j} \Theta_{i,j}\sigma_i^x\sigma_j^x\right),
\label{eq:unitary}
\end{equation}
where $\phi_j = \sum_{m} \left(\alpha_{j,m} a_m^\dag - \alpha_{j,m}^*a_m\right)$ and
\begin{equation}
\alpha_{j,m} = -\frac{i}{\hbar} \eta_{j,m} \int_0^\tau f_j(t)e^{i\omega_m t}dt
\label{eq:alpha}
\end{equation}
\begin{align}
\Theta_{i,j} = &\frac{1}{\hbar^2} \sum_m \eta_{i,m} \eta_{j,m} \int_0^\tau dt_1 \int_0^{t_1} dt_2 \sin\omega_m(t_1-t_2) \times\nonumber\\
& \qquad\qquad\qquad\left[f_i(t_1)f_j(t_2)+f_j(t_1)f_i(t_2)\right].
\label{eq:theta}
\end{align}

For a high-fidelity realization of the entangling gate, we drive ions $i$ and $j$, set $\Theta_{i,j}=\pm\pi/4$, and minimize the $\alpha_{j,m}$ terms by suitable amplitude, phase, or frequency modulation \cite{zhu_trapped_2006,green_phase-modulated_2015,leung_robust_2018,choi_optimal_2014,green_phase-modulated_2015,lin_large-scale_2009}. The average gate infidelity intrinsic in this gate design from residual entanglement with the phonon modes is approximately \cite{wu_noise_2018}
\begin{equation}
\delta F = \frac{4}{5}\sum_{j,m} |\alpha_{j,m}|^2 (2\bar{n}_m + 1)
\label{eq:intrinsic_fidelity}
\end{equation}
where $\bar{n}_m$ is the average phonon occupancy for mode $m$.


\section{Parallel Gates Theory and Simulation}
In this section we demonstrate how the design of the entangling gates in a large ion crystal is insensitive to the distant ions and how it can be parallelized, providing a new avenue towards connectivity and scalability. Parallel entangling gates have been discussed in Ref. \cite{figgatt_parallel_2018,lu_scalable_2019} for arbitrary ion pairs, but designing such parallel gates may require significant experimental resources for long ion chains.
Here, we tackle the problem through a different approach: by considering entangling gates on only nearby ions, distant gates
can be parallelized without any overhead in gate design compared with that on a small crystal.
Later we will show that this can still lead to an efficient realization of a complex quantum gate if it can be decomposed
into a few layers of local entangling gates.

We first consider an infinite ion chain with uniform inter-ion spacing $d$, in order to simplify the derivation; although neither the gate design nor its parallelization relies on these assumptions. For realistic ion crystals, the spacing is never uniform, but a good approximation can be achieved by adding non-quadratic terms to the electric potential along the axial direction \cite{lin_large-scale_2009}. Note that for an infinite chain, a continuous spectrum of transverse modes takes the form of sinusoidal travelling waves. The modes can be characterized by a wavenumber $\kappa$ and ion position coordinate $z_j$, with the mode vector $b_{j,\kappa}\propto \exp(i \kappa z_j)$ and the frequency
\begin{equation}
\omega_\kappa \approx\omega_1 \left\{1-\epsilon
\left[\zeta(3)-S(\kappa d)\right]\right\},
\end{equation}
where $\omega_1$ is the transverse center-of-mass mode frequency and $\epsilon=e^2/(4\pi\epsilon_0 d^3 m\omega_1^2) \sim \sqrt{(logN)/N}$ is a small parameter due to the anisotropy of the linear ionic crystal
and describes the narrowness of the entire transverse spectrum \cite{dubin_theory_1993,Schiffer_phase_1993}.
$\zeta(3) \equiv\sum _{j=1}^{\infty}1/j^{3}\approx 1.202$ and $S(x)\equiv \sum_{j=1}^\infty (\cos jx) / j^3$. More details can be found in Appendix~A.

\begin{figure}[htbp]
  \centering
  \includegraphics[width=89mm]{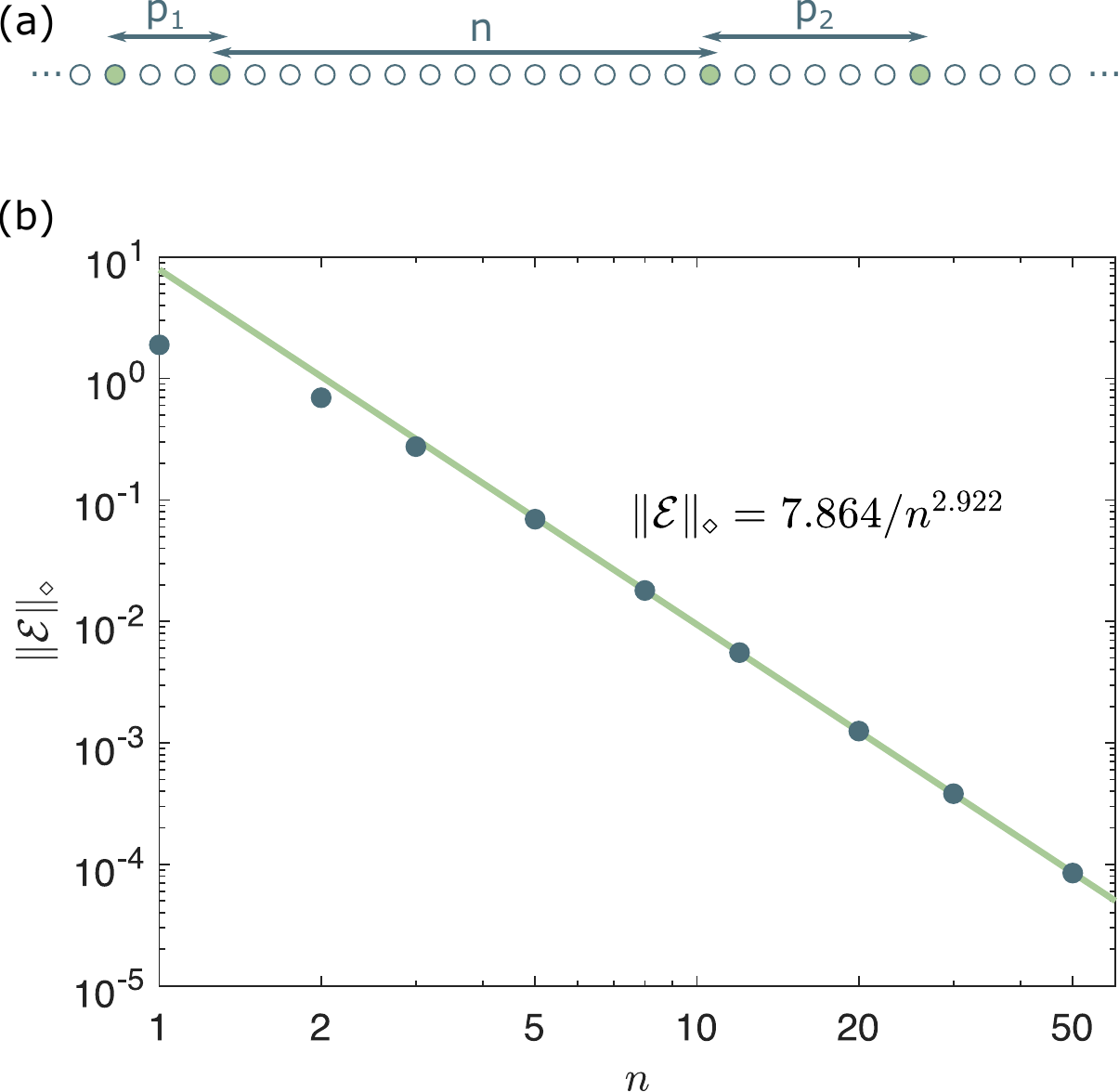}\\
  \caption{(a) Schematic of a long ion chain with two pairs selected to represent a simultaneous entangling gate as described in the main body. Note that the minimal inter-gate distance between pairs is $n$, and $p$ is the intra-gate distance between ion pairs.
  (b) Log-log plot for crosstalk error $\|\mathcal{E}\|_\diamond$ vs. gate distance $n$ on a 100-ion chain.
  The gate is designed for a nearest-neighbor pair of ions with $\tau=50\,\mu$s, $n_{\mathrm{seg}}=6$ segments, and $\mu=1.016\omega_1$. The green line is fitted from the last five data points.
  More information can be found in Appendix~A}\label{fig:cross_scaling}
\end{figure}

With the motional mode spectrum well characterized, we now consider implementing two gates between ions $i_1$ and $j_1$ and between $i_2$ and $j_2$ with respective intra-gate distances $p_1=|i_1-j_1|$, $p_2=|i_2-j_2|$, and $p=\max(p_1,p_2)$ (see Fig.(\ref{fig:cross_scaling})). Illuminating all four ions simultaneously creates the ideal evolution operator of the form $\exp(\pm i\pi/4 \sigma_{i_1}^x\sigma_{j_1}^x) \exp(\pm i\pi/4 \sigma_{i_2}^x\sigma_{j_2}^x)$, as well as unwanted gate errors. 
The intrinsic gate errors from Eq.~(\ref{eq:intrinsic_fidelity}) add cumulatively for each ion and mode combination, and can be minimized
%
by engineering high-fidelity designs for the two entangling gates individually \cite{zhu_arbitrary-speed_2006, lin_large-scale_2009, choi_optimal_2014, green_phase-modulated_2015, debnath_demonstration_2016, leung_robust_2018}.
From Eq.~(\ref{eq:theta}), we can see that the entanglement parameter $\Theta_{i_1 ,j_1}$ only depends on the laser forces on ions $i_1$ and $j_1$ and is not affected by the laser drive on ions $i_2$ and $j_2$, so its value can be set to $\pm \pi/4$ by design; the same argument holds for $\Theta_{i_2,j_2}$. However, there are four additional crosstalk terms between the inter-gate ion pairs. Because all these entangling operators commute with each other, the nominal evolution of pairs $\{i_1, j_1\}$ and $\{i_2, j_2\}$ (including the intrinsic errors from the gate design) are accompanied by an additional unitary operator $\mathcal{C}(\rho)=V\rho V^\dag$ with
\begin{equation}
V = \exp\left[i \sum_{r=i_1,j_1}\sum_{s=i_2,j_2}\Theta_{r,s}\sigma_{r}^x\sigma_{s}^x\right].
\end{equation}
We characterize the net error in the parallel gates using the diamond norm metric because it bounds the error rate \cite{sanders_bounding_2015} and includes coherent errors \cite{kueng_comparing_2016}. The diamond norm of the error $\mathcal{E}\equiv \mathcal{C}-\mathcal{I}$ is bounded by crosstalk terms as 
$\frac{1}{2}\|\mathcal{E}\|_\diamond \le |\Theta_{i_1, i_2}| + |\Theta_{i_1, j_2}| + |\Theta_{j_1, i_2}| + |\Theta_{j_1, j_2}|$
\cite{aharonov_quantum_1998}, and we say that two entangling gates can be parallelized if $\|\mathcal{E}\|_\diamond \ll 1$.

We next quantify crosstalk errors in the context of two parallel gates by considering how the entanglement parameter $\Theta_{i,j}$ scales with $n$, the minimal inter-gate distance between ion pairs.
While the derivation of Eq.~(\ref{eq:theta}) assumes real mode vectors $b_{j,k}$, complex mode vectors 
considered in this infinite chain limit lead to a similar expression that can be shown to decay as $1/n^3$ 
(see Appendix~A for more details). 
Therefore, when two gates are applied simultaneously, the error due to the parallelization of the gates is:
\begin{equation}
\|\mathcal{E}\|_\diamond\lesssim 2\pi(p/n)^3
\end{equation}

This analytical result
shows that simultaneous gates will create significant levels of crosstalk entanglement if $p\sim n$, but this unwanted entanglement falls off with the cube of the distance between the two pairs.
This is supported by our numerical simulations on a finite length crystal presented in Fig.~\ref{fig:cross_scaling}b without the approximations made in the derivation. We consider a finite chain of $N=100$ ions with a uniform spacing $d=8\,\mu$m and a highest transverse normal mode frequency $\omega_1=2\pi\times 3\,$MHz. A gate is designed for a nearest-neighbor ion pair with $n_{\mathrm{seg}}=6$ amplitude segments, gate time $\tau=50\,\mu$s, and detuning from qubit resonance $\mu = 1.016\omega_1$. This detuning is slightly higher than the highest (in phase) normal mode sideband. (As we show in Appendix~A, the gate design is not sensitive to the position of the ion pairs inside the chain.) We apply the Ising gate sequence onto ions $i_1=20$ and $j_1=21$ in the ion chain as well as ions $i_2=22,\,\cdots,\,71$ and $j_2=i_2+1$ (we use the central part of the chain to avoid boundary effects on the scaling). At every distance between the pairs of ions, $n=|i_2-j_1|$, we calculate the error due to the crosstalk terms; the $1/n^3$ scaling is clear for large $n$, see Fig.~\ref{fig:cross_scaling}b. Note that here we assume the same gate design for the two parallel gates only for the convenience of evaluating $\Theta_{i,j}$ [Eq.~(\ref{eq:theta})] numerically. The derivation of the $1/n^3$ scaling does not rely on this assumption and holds for parallelizing two different entangling gates.

This result can directly be applied to parallelize multiple gates. Suppose we want to build a quantum circuit with Ising gates between all possible ion pairs whose intra-gate distance is less than or equal to $p$. We can divide these gates into $\mathcal{O}[p(n+p)]$ layers in such a way that the gates are separated by a distance of at least $n$ in each layer. In this way, the error per gate in each layer will be $\mathcal{O}[(p/n)^3]$. Hence for a given error rate per gate $\epsilon$, the number of required layers will be independent of the size of the crystal, a criterion for scalability.
In Appendix~A we also show that the same scaling law holds for a 2D hexagonal lattice, which suggests that this scalability may be universal and can be applied to more general ion crystals.

\begin{figure}[htbp]
  \centering
  \includegraphics[width=0.8\linewidth]{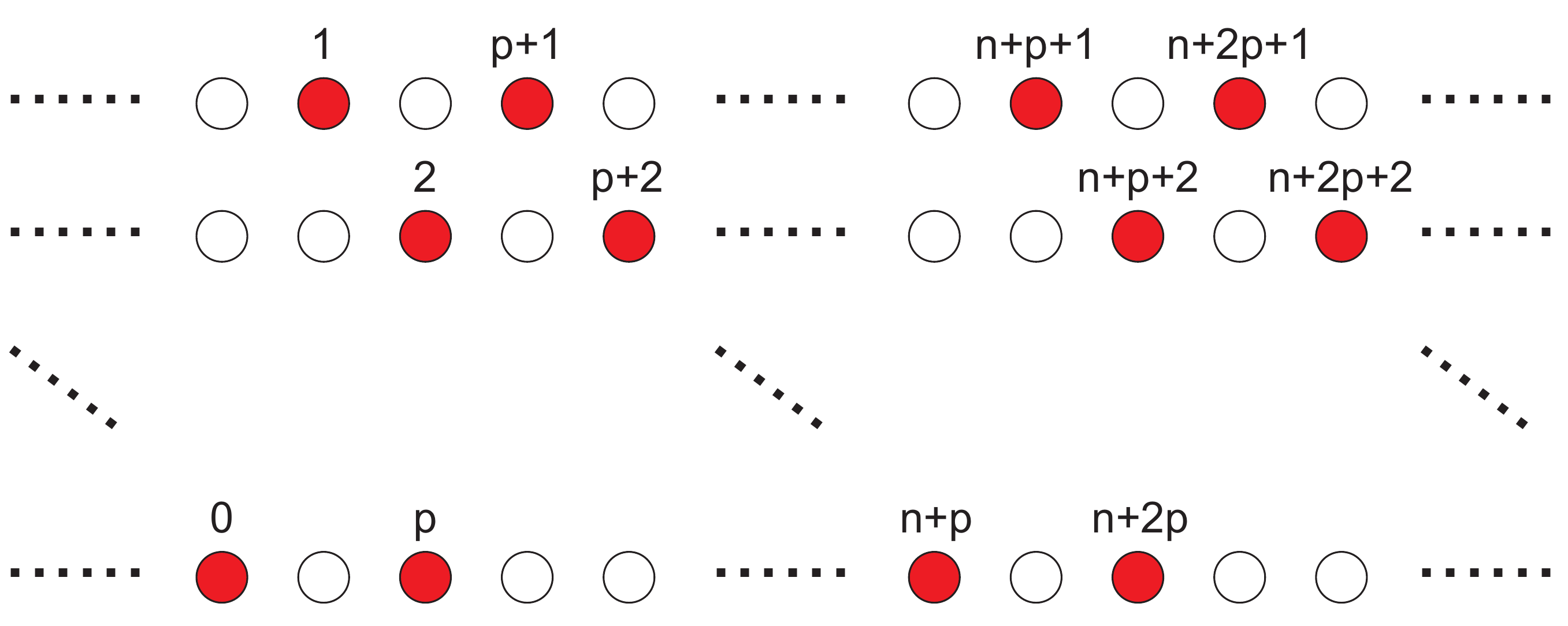}
  \caption{Partial illustration of the scheme to parallelize gates. $(n+p)$ layers are needed for all possible pairs of ions at an intra-gate distance of $p$ while maintaining an inter-gate distance of $n$ between any two pairs.
  For all possible gates whose distance is less than or equal to $p$, we need $\mathcal{O}[p(n+p)]$ layers in total.}
\end{figure}

\begin{figure*}[htbp]
\centering
\includegraphics[width=\textwidth]{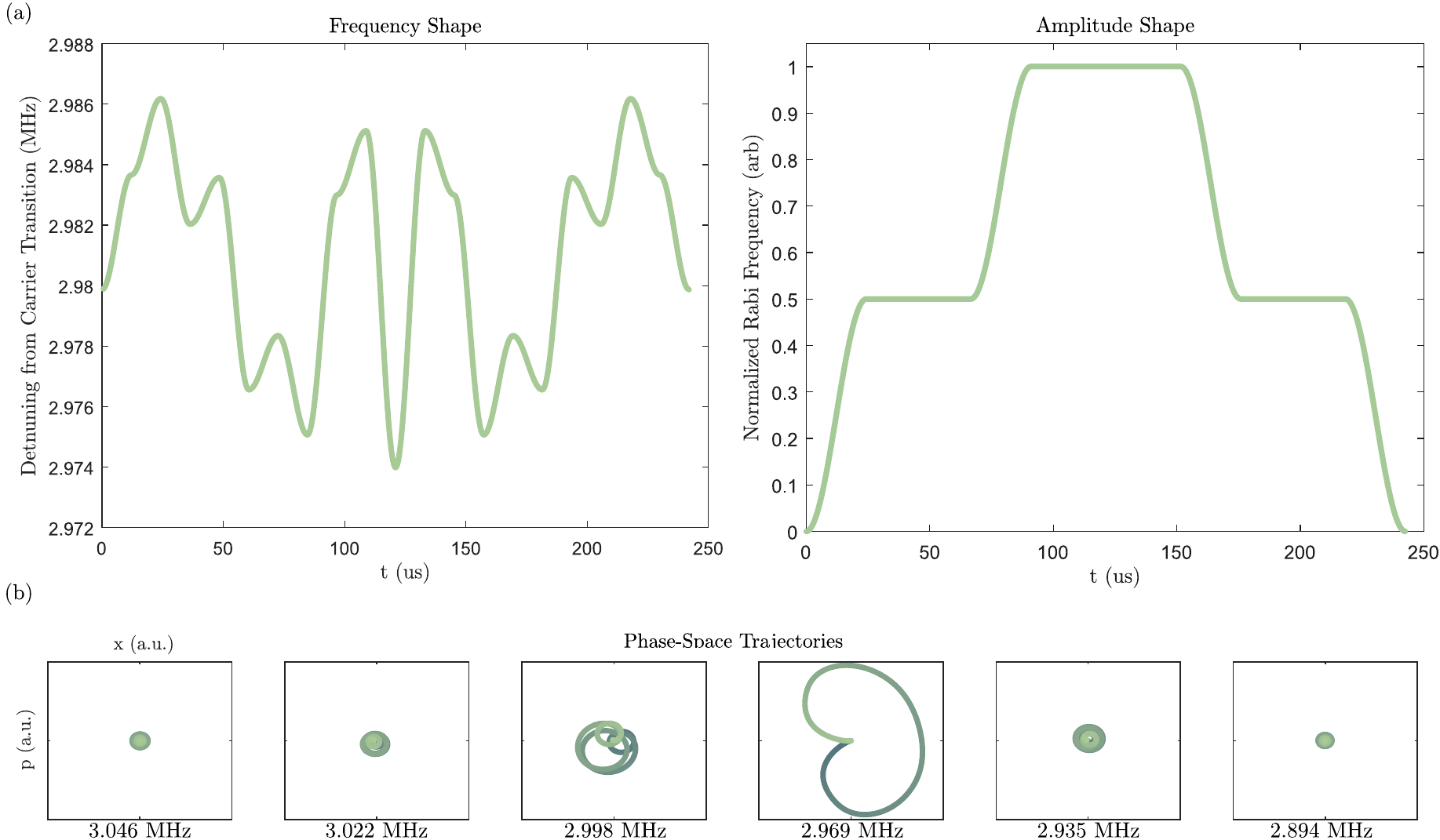}
\caption{(a) Frequency and amplitude modulation are plotted for the experimental gate performed on a chain of 17 ions. Phase-space trajectories are closed by changing the frequency modulation, while the amplitude modulation is designed to prevent unwanted Fourier components in the qubit drive. 
(b)
Trajectories of ions through phase space during the gate depicted in (a). Each phonon mode is labeled by its frequency under its respective phase space plot. The majority of entanglement comes from the geometric phase accumulated when coupling with the 2-5 highest energy phonons. The rest of the phonons are relatively unexcited and only the highest energy 6 phonon modes are plotted, the remaining are included in the appendix for completeness. The Lamb-Dicke (LD) parameter used for all these plots is based on the common-mode LD parameter.
}
\label{fig:GatePSTraj}
\end{figure*}

Finally, insensitivity of an entangling gate to operations on other distant ions suggests that we can also neglect these ions when designing the gate. Therefore we only need to consider a small number of ions and their oscillation modes in Eq.(\ref{eq:alpha}) and Eq.(\ref{eq:theta}). This is supported by an numerical example in Appendix~A and the experiment below, and permits a scalable method for designing gates in a large crystal where full connectivity may be impractical due to the falloff of the Coulomb interaction with distance. 

\section{Gates on 17-Ion Chains}
Next, we demonstrate a component of this scalability by operating high-fidelity gates between ions in a long chain with relatively simple pulse sequences that are designed while ignoring far-detuned motional modes.  We demonstrate Ising gates on pairs of ions in a system of $17$  tightly-confined \Yb qubits. The ions are confined in an rf Paul trap with a transverse trapping frequency of $\omega_1 = 2\pi\times 3.04\:$ MHz. The qubit states are defined by two hyperfine-split states in the $^2S_{1/2}$ ground-level manifold as $\ket{0} = \ket{F=0,m_F=0},~\ket{1} = \ket{F=1,m_F=0}$. The qubit splitting is $2\pi\times12.642821\:$GHz and is nearly magnetic field insensitive \cite{olmschenk_manipulation_2007}. We implement coherent operations on the ions using counter-propagating optical Raman beams at 355 nm that create a beat-note resonant with the qubit \cite{islam_beat_2014}. One Raman beam illuminates the entire chain, while the other is split into individual beams that are each controlled by unique channels of a multi-channel acousto-optic modulator (AOM) and are then focused onto single ions \cite{debnath_demonstration_2016}. In this way, we effect individual addressing of ion numbers 5,7,9,11, and 13 in the chain with full amplitude, frequency, and phase control. For this experiment, we designed the harmonic confinement in the axial direction such that the ion spacing for the middle nine ions is roughly  $2.5\: \mu$m. This allows us to align the five individual laser beams, which are $5\: \mu$m apart, onto the five ions. These ions are also matched onto individual channels of a multi-channel photo-multiplier tube (PMT) to accomplish individual detection. 

Using experimentally measured transverse phonon mode frequencies for the $6$ highest-energy modes, amplitude and frequency modulated gates are calculated for chains of $17$ ions using the methods described in Ref. \cite{leung_entangling_2018}. Only a small subset of modes is used to calculate the gates, which tests the ability of the solution generator to produce high-fidelity solutions with limited knowledge of the system.  The modulation is graphically depicted in Fig. \ref{fig:GatePSTraj}. By allowing the drive frequency to vary with time, the several degrees of freedom needed to satisfy the requirements for the Ising gate are fulfilled. Therefore, the dependence on $\omega_m$ in Eq. \ref{eq:alpha} becomes time-dependent, as described in detail in Ref. ~\cite{leung_entangling_2018}.



During the Ising gate, the lasers transfer momentum into the Coulomb crystal. The excited phonon's trajectory through phase space is determined in part by the detuning of the laser drive, as seen in Eq. \ref{eq:alpha}. Since the ion spacing is much lower than previous experiments \cite{debnath_demonstration_2016,landsman_verified_2019,figgatt_complete_2017}, the spectral gap between transverse mode frequencies is nearly twice as large: $25$ kHz on average. Therefore, the nominal frequency of the calculated gate is far-detuned from most of the modes even though it sits inside the mode spectrum. This can be seen by the phase-space plots of the motional excitation, $\alpha_{j,m}$ for one of the selected ions as shown in Fig \ref{fig:GatePSTraj}, which are already negligibly small for the sixth phonon mode due to the large detuning. The lower frequency phonon modes have even more tightly confined trajectories.

The Ising gate is performed between ion pairs \{5,13\} and \{7,9\} with fidelities of $97(1)\%$ and $95(1)\%$, respectively, compared with a theoretical maximum fidelity of $1 - 4.7 \times 10^{-4}$ (fundamental limits of spontaneous emission are much lower). 
These results are corrected for state preparation and measurement errors of $\approx1.8\%$. We calculate the gate fidelity from population measurements and the parity oscillation amplitude with the phase of a global qubit rotation ~\cite{choi_optimal_2014, sackett_experimental_2000}. 
\begin{figure}
\centering
\includegraphics[width = 89 mm]{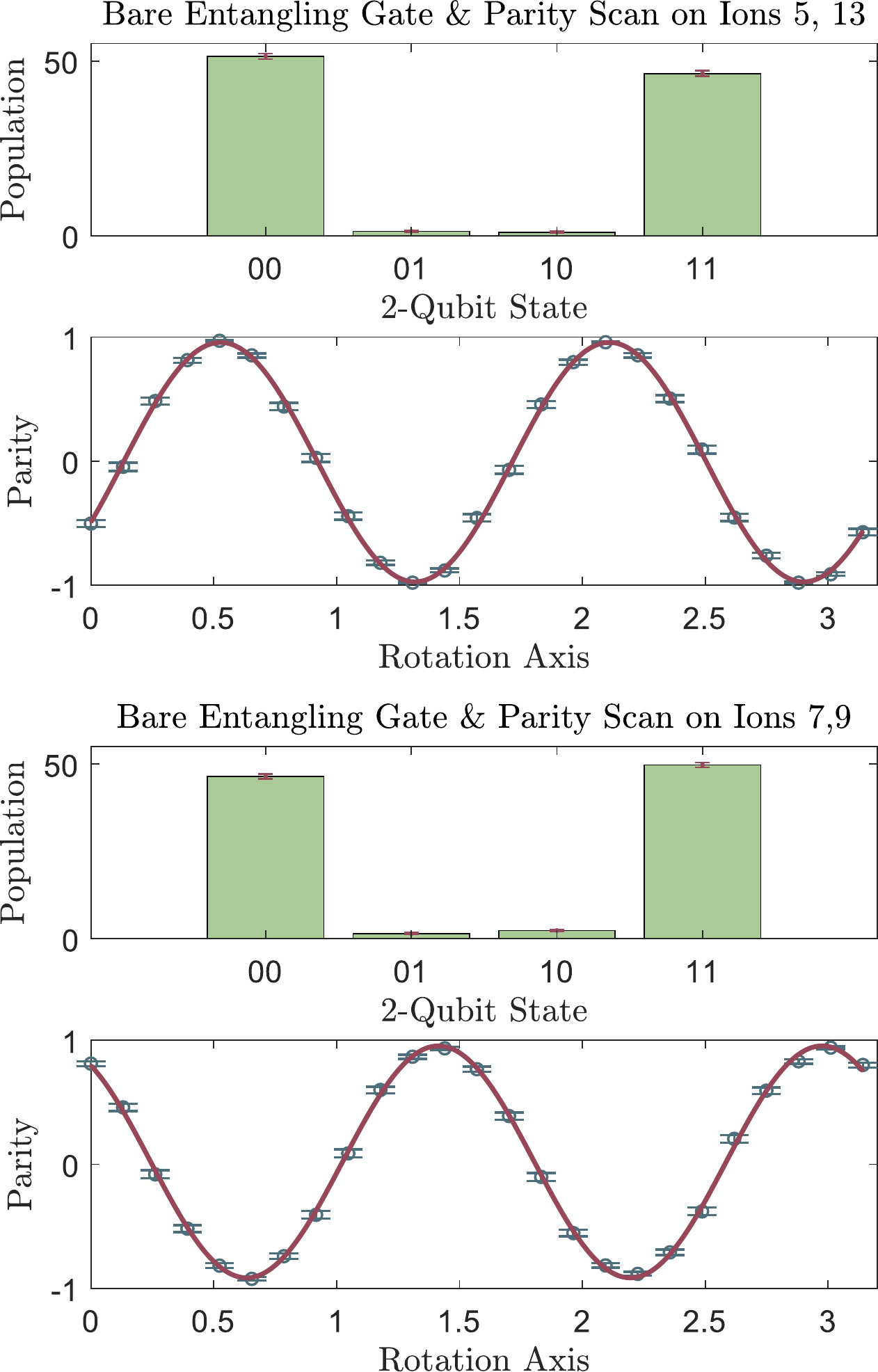}
\label{fig:GateResults}
\caption{Bare gate and parity scan data are plotted for two qubit gates on ions \{5,13\} and \{7,9\} in a 17 ion chain. The fidelities are $97(1)\%$ and $95(1)\%$, respectively. The amplitude was deduced from a least-squares fit, and error bars are statistical errors based on a binomial distribution of individual qubit states.}
\end{figure}


\section{Outlook}
In the limit of large chains, where all ions are coupled to all motional modes with similar strength, we theoretically show the ability to entangle multiple short-distance ion pairs simultaneously with minimal crosstalk entanglement. In contrast, when only a minority of motional modes are coupled, it is more useful to entangle arbitrary pairs of ions over long distances, as we have demonstrated experimentally.
Both results suggest that adding ions to a trap need not affect gate fidelity. 
The gates performed on strings of 17 ions are comparable to those obtained on the very same hardware with few ions in the trap; with only 5 ions, fidelities of roughly $98\%$ were recorded in Ref. \cite{debnath_demonstration_2016}. Similarly, the gate used for the simulations in the appendix does not change if additional ions are added into the trap. These results point to the inherent scalability of performing entangling gates on trapped ion qubits. 

\section{Acknowledgements}
We thank S. Debnath for insightful conversations. This work is supported in part by the ARO through the IARPA LogiQ program, the AFOSR MURI on Quantum Measurement and Verification, 
the NSF STAQ Practical Fully-Connected Quantum Computer program, 
the ARO MURI on Modular Quantum Circuits, 
the DOE ASCR, BES quantum computing programs, 
and the NSF Physics Frontier Center at JQI.

\bibliography{LongChainBib}

\newpage
\subsection{Appendix A: Theoretical Details on Parallel Gates}\label{appendixA}
\subsubsection{1D Uniform Chain}
Consider an infinite ion chain along the $z$ axis with uniform spacing $d$. The ions can be labelled by integers with ion $j$ located at $z_j=jd$.
Suppose the trapping potential along the $x$ direction is harmonic with trapping frequency $\omega_1$, then the transverse oscillation modes in this direction can be expressed as travelling waves
\begin{equation}
b_{j,\kappa} = \frac{1}{\sqrt{N}} e^{i\kappa z_j}
\end{equation}
where $N$ is the number of ions and we will later take the limit $N\to \infty$. The mode frequencies are
\begin{align}
\omega_\kappa =& \omega_1 \sqrt{1-\frac{e^2}{4\pi\epsilon_0 d^3 m\omega_1^2}
\sum_{j\ne 0} \frac{1-\cos (j\kappa d)}{|j|^3}}\nonumber\\
=& \omega_1 \sqrt{1-\frac{e^2}{4\pi\epsilon_0 d^3 m\omega_1^2}
2\left[\zeta(3)-S(\kappa d)\right]}
\end{align}
where $m$ and $e$ are the mass and the charge of the ion, $\zeta(3) \equiv\sum _{j=1}^{\infty}1/j^{3}\approx 1.202$ is the Riemann zeta function and $S(x)\equiv \sum_{j=1}^\infty (\cos jx) / j^3$.

Define $\epsilon=e^2/(4\pi\epsilon_0 d^3 m\omega_1^2)$, which is typically small. We get
\begin{equation}
\omega_\kappa \approx\omega_1 \left\{1-\epsilon
\left[\zeta(3)-S(\kappa d)\right]\right\}.
\end{equation}

Where $\omega_1$ is the transverse center-of-mass mode frequency.
As is mentioned in the main text, our expression for the entanglement parameter Eq.~(\ref{eq:theta}) needs to be modified for the complex mode vectors
\begin{widetext}
\begin{align}
\Theta_{i,j} =& \frac{1}{N}\sum_{\kappa} g_\kappa^2 \int_0^\tau dt_1
\int_0^{t_1}dt_2
\Big\{\left[f_i(t_1) f_j(t_2) + f_j(t_1) f_i(t_2)\right]\sin\omega_\kappa (t_1-t_2)\cos \kappa(z_j-z_i)\nonumber\\
& \qquad\qquad\qquad + \left[f_i(t_1) f_j(t_2) - f_j(t_1) f_i(t_2)\right]\cos\omega_\kappa (t_1-t_2)\sin \kappa(z_j-z_i)\Big\}
\end{align}
\end{widetext}
Here we separate the mode vector $b_{j,\kappa}$ from $\eta_{j,\kappa}$ appearing in Eq.~(\ref{eq:theta}) as $\eta_{j,\kappa}=g_\kappa b_{j,\kappa}$ with $g_{\kappa}=\sqrt{\hbar k^2 /(2m\omega_\kappa)}$, since we already have the analytical expression for the mode vectors.

We further replace $\frac{1}{N}\sum_\kappa$ with $\frac{d}{2\pi}\int d\kappa$ and change the order of integration. Without loss of generality we assume $j>i$ and define $z_j-z_i=nd$. Because we are interested in the scaling with $n$,
we only need to evaluate
\begin{equation}
\label{eq:term_1}
\int_{-\pi}^{\pi} d(\kappa d) g_\kappa^2 \sin \omega_\kappa(t_1-t_2) \cos n\kappa d
\end{equation}
and
\begin{equation}
\label{eq:term_2}
\int_{-\pi}^{\pi} d(\kappa d) g_\kappa^2 \cos \omega_\kappa(t_1-t_2) \sin n\kappa d
\end{equation}
while the rest of $\Theta_{i,j}$ only depends on the laser sequence on the ions but not on their distance.
Here we calculate the first term as an example; the second one can be treated
similarly.

Let us define $\phi = \omega_1(t_1-t_2)[1-\epsilon \zeta(3)]$ and
$\lambda=\epsilon\omega_1(t_1-t_2)$. Lamb-Dicke parameter $g_\kappa$ depends on $\kappa$ as $g_\kappa^2=g_0^2 \omega_1/\omega_\kappa$, where $g_0$ is a constant independent of $\kappa$. Hence we can approximate $g_\kappa$ by $g_0$ with an error of $\epsilon$.
Then we have
\begin{align}
&\int_{-\pi}^{\pi} d(\kappa d) \sin \omega_\kappa(t_1-t_2) \cos n \kappa d \nonumber\\
=&\int_{-\pi}^{\pi} dx \left\{\sin \phi \cos [\lambda S(x)] + \cos \phi \sin [\lambda S(x)] \right\} \cos nx
\label{eq:term_1_all}
\end{align}

Again we consider the first term as an example, while the second term can be calculated
in the same way.  Plugging in the series expansion forms $\cos x = \sum_{\alpha=0}^\infty (-1)^\alpha x^{2\alpha} / (2\alpha)!$ and $S(x)= \sum_{\beta=1}^\infty \cos \beta x / \beta^3$, we get
\begin{align}
&\int_{-\pi}^{\pi} dx \cos [\lambda S(x)]
\cos nx \nonumber\\
=&\sum_{\alpha=0}^\infty \frac{(-1)^\alpha}{(2\alpha)!}\int_{-\pi}^{\pi} dx \left(\lambda \sum_{\beta=1}^\infty \frac{\cos \beta x}{\beta^3}\right)^{2\alpha} \cos nx.
\label{eq:term_1_1}
\end{align}

Now we argue that this expression has a scaling of $1/n^3$. First, note that for a given $\alpha$, the integrand can be expanded into a series
\begin{equation}
\label{eq:analytic_expansion}
\sum_{\{\beta_j\}} \lambda^{2\alpha} \cos nx \prod_{j=1}^{2\alpha} \frac{\cos \beta_j x}{\beta_j^3},
\end{equation}
with each term being a product of $(2\alpha+1)$ cosine functions and the integration over their common period; therefore the integral is nonzero only if a resonance condition $n\pm \beta_1 \pm \beta_2 \pm \cdots \pm \beta_{2\alpha}=0$ is satisfied, and in such cases it can be loosely bounded by
\begin{equation}
\left|\int_{-\pi}^{\pi} dx \cos nx \prod_{j=1}^{2\alpha}\cos \beta_j x \right| \le \int_{-\pi}^{\pi} dx \cdot 1 = 2\pi
\end{equation}

Furthermore, the dominant term of Eq.~(\ref{eq:analytic_expansion}) should contain one and only one $\beta_j$ of the order $O(n)$; all the other $(2\alpha-1)$ terms of $\beta_j$'s must be bounded by a constant, say, $C/2$, otherwise their coefficients will decay faster than $1/n^3$.
Admittedly, for $\alpha$ of the order $O(n)$ all the $\beta_j$'s can be of the order $O(1)$, but then
the $1/(2\alpha)!$ factor in Eq.~(\ref{eq:term_1_1}) itself decays faster than $1/n^3$.

There are in total $2\alpha C^{2\alpha-1}$ such terms in Eq.~(\ref{eq:analytic_expansion}): first we choose which of the $2\alpha$ terms of $\beta_j$'s takes the order of $O(n)$; then we assign the other $(2\alpha-1)$ terms of $\beta_j$'s within $[1,C/2]$ and choose their signs in the resonance condition to be positive or negative; once we determine these values, the last one is automatically fixed from the resonance condition.
Therefore finally Eq.~(\ref{eq:term_1_1}) is bounded by
\begin{equation}
2\pi\sum_{\alpha=1}^{\infty} \frac{2\alpha C^{2\alpha-1}\lambda^{2\alpha}}{(2\alpha)!} \frac{1}{n^3} = \frac{2\pi\lambda\sinh \lambda C}{n^3}\propto \frac{1}{n^3}
\end{equation}
Note that $\lambda$ actually depends on $t_1$ and $t_2$, but here we only consider the scaling
with respect to $n$. Nevertheless, for typical parameters, $|\lambda|\le \epsilon \omega_1 \tau$
is limited to the order of $O(1)$.

Similar argument applies to the other terms in Eq.~(\ref{eq:term_2}) and Eq.~(\ref{eq:term_1_all}), therefore we conclude that the entanglement parameter $\Theta_{i,j}$ decays with the ion spacing $n=|i-j|$ as $1/n^3$. We also show some numerical examples to verify this scaling at the end of this appendix.

\subsubsection{2D Hexagonal Lattice}
Now we generalize our results to 2D. Due to the Coulomb interaction between the ions, a 2D ion crystal usually approximates a hexagonal lattice. Therefore here we consider a hexagonal lattice with translational symmetry, although the same analysis can also be applied to other types of 2D lattices.

Let the crystal lie on the $x$-$y$ plane. It can be described by its lattice vectors
\begin{equation}
\boldsymbol{a}_1 = d\left(1, 0, 0\right), \qquad \boldsymbol{a}_2 = d\left(\frac{1}{2}, \frac{\sqrt{3}}{2}, 0\right)
\end{equation}
with the corresponding reciprocal vectors
\begin{equation}
\boldsymbol{b}_1 = \left(1, -\frac{1}{\sqrt{3}}, 0\right), \qquad \boldsymbol{b}_2 = \left(0, \frac{2}{\sqrt{3}}, 0\right)
\end{equation}

The transverse oscillation modes in the $z$ direction can be used for the entangling gates \cite{wang_quantum_2015}. In this case the mode vector for a wave vector $\boldsymbol{\kappa}=\kappa_1 \boldsymbol{b}_1 + \kappa_2 \boldsymbol{b}_2$ ($\kappa_1, \kappa_2 \in (-\pi/d,\pi/d]$) is given by
\begin{equation}
z_{\alpha\beta}^{\boldsymbol{\kappa}} \propto \exp \left[ i(\alpha \kappa_1 d + \beta \kappa_2 d) \right]
\end{equation}
for the ion at position $\boldsymbol{r}_{\alpha\beta} = \alpha \boldsymbol{a}_1 +\beta \boldsymbol{a}_2$.

The corresponding mode frequency is
\begin{align}
\omega_{\boldsymbol{\kappa}} =& \omega_z \sqrt{1 - \frac{e^2}{4\pi\epsilon_0 d^3 m\omega_z^2} {\sum_{(\alpha,\beta)}}' \frac{1-\cos \boldsymbol{\kappa}\cdot\boldsymbol{r}_{\alpha\beta}}{|\alpha \boldsymbol{a}_1 +\beta \boldsymbol{a}_2|^3}}\nonumber\\
=& \omega_z \sqrt{1 - \frac{e^2}{4\pi\epsilon_0 d^3 m\omega_z^2} {\sum_{(\alpha,\beta)}}' \frac{1-\cos (\alpha \kappa_1 d + \beta \kappa_2 d)}{(\alpha^2+\beta^2+\alpha\beta)^{3/2}}}\nonumber\\
\approx& \omega_z \left[1 - \frac{e^2}{4\pi\epsilon_0 d^3 m\omega_z^2} {\sum_{(\alpha,\beta)}}' \frac{1-\cos (\alpha \kappa_1 d + \beta \kappa_2 d)}{2(\alpha^2+\beta^2+\alpha\beta)^{3/2}}\right]
\end{align}
where ${\sum}'$ means that the two indices for summation cannot both be 0 and $\omega_z$ is the secular frequency in the z direction.

Following the same derivation as before, to study the scaling of the crosstalk error versus the distance on the lattice represented by $n\boldsymbol{a}_1+m\boldsymbol{a}_2$, we need to evaluate some expressions like
\begin{align}
&\sum_{\alpha=0}^\infty \frac{(-1)^\alpha}{(2\alpha)!}\int_{-\pi}^{\pi} dx \int_{-\pi}^{\pi} dy \left[\frac{\lambda}{2} {\sum_{{\beta,\gamma}}}' \frac{\cos (\beta x + \gamma y)}{(\beta^2+\gamma^2+\beta\gamma)^{3/2}}\right]^{2\alpha}\nonumber\\
& \qquad\qquad\qquad\times\cos (nx + my)
\end{align}
Again for a given $\alpha$ we can expand the integrand into series:
\begin{equation}
\sum_{\{\beta_j,\gamma_j\}} \left(\frac{\lambda}{2}\right)^{2\alpha} \cos (nx + my) \prod_{j=1}^{2\alpha} \frac{\cos (\beta_j x + \gamma_j y)}{(\beta_j^2+\gamma_j^2+\beta_j\gamma_j)^{3/2}}
\end{equation}
and we get two resonance conditions $n\pm \beta_1 \pm \cdots \pm \beta_{2\alpha}=0$ and $m\pm \gamma_1 \pm \cdots \pm \gamma_{2\alpha}=0$, both of which need to be satisfied for a nonzero integral.

Without loss of generality, we can assume $|n|\ge|m|$. Consider two different cases: (1) only $|n|$ goes to infinity and $|m|$ stays constant. Then we go back to the previous 1D case and the coefficient decays as $1/n^3$. (2) Both $|n|$ and $|m|$ go to infinity. Then again we argue that
for any given $\alpha$, the $2\alpha$ terms of $\{\beta_j,\,\gamma_j\}$ can only have one term of the order
$O(|n|)$ and $O(|m|)$; all the other $\beta_j$'s and $\gamma_j$'s need to be
bounded by constant. Otherwise the coefficient for their product will decay faster than $1/(n^2+m^2+nm)^{3/2}$. If we count the number of such terms and add all of them together, we will get a scaling of $1/(n^2+m^2+nm)^{3/2}$, that is, a cubic decay with the distance.

\subsubsection{Numerical Results}
\begin{figure}[htbp]
\centering
\subfigure[]{\includegraphics[width=40mm]{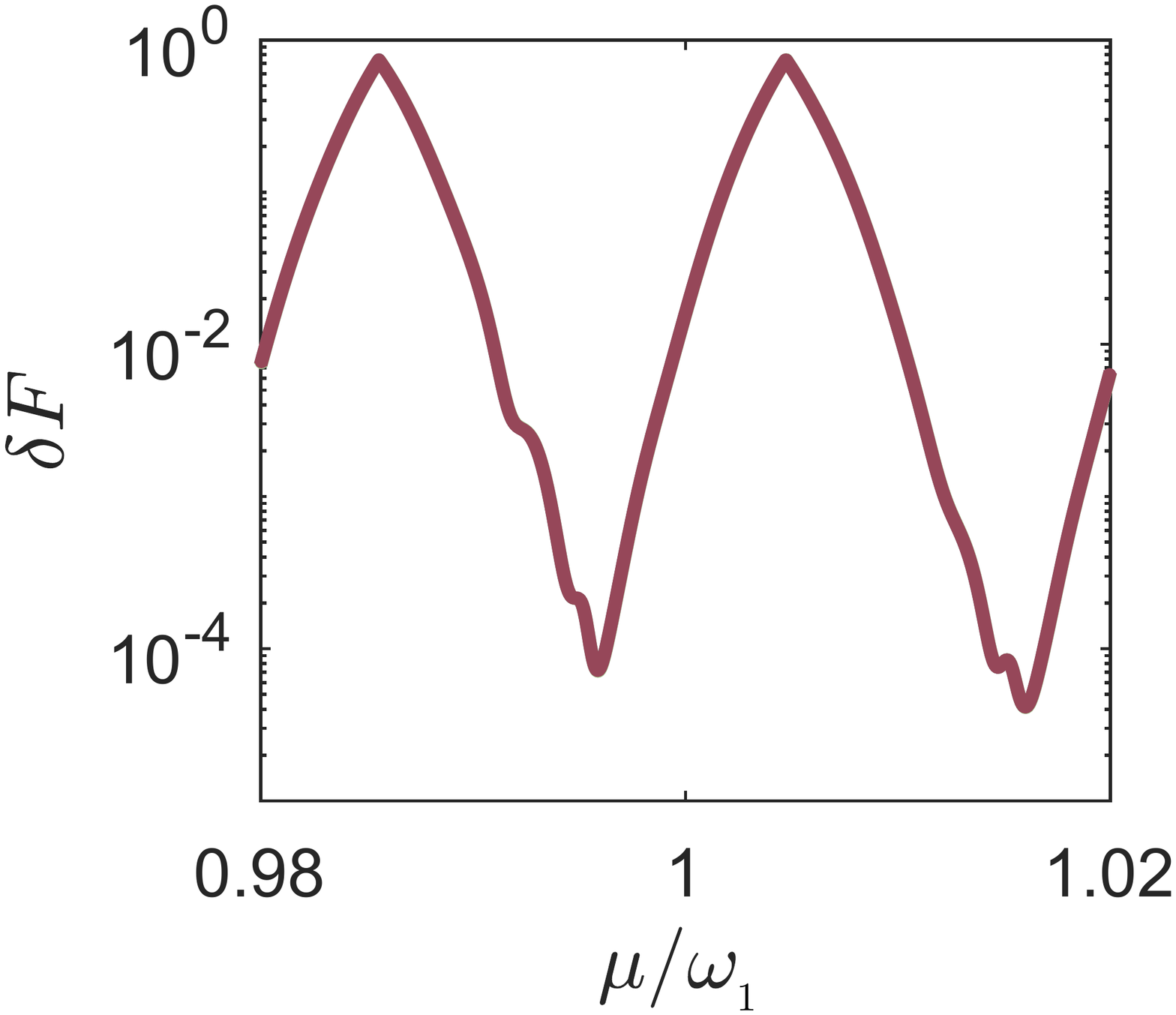}}
\subfigure[]{\includegraphics[width=40mm]{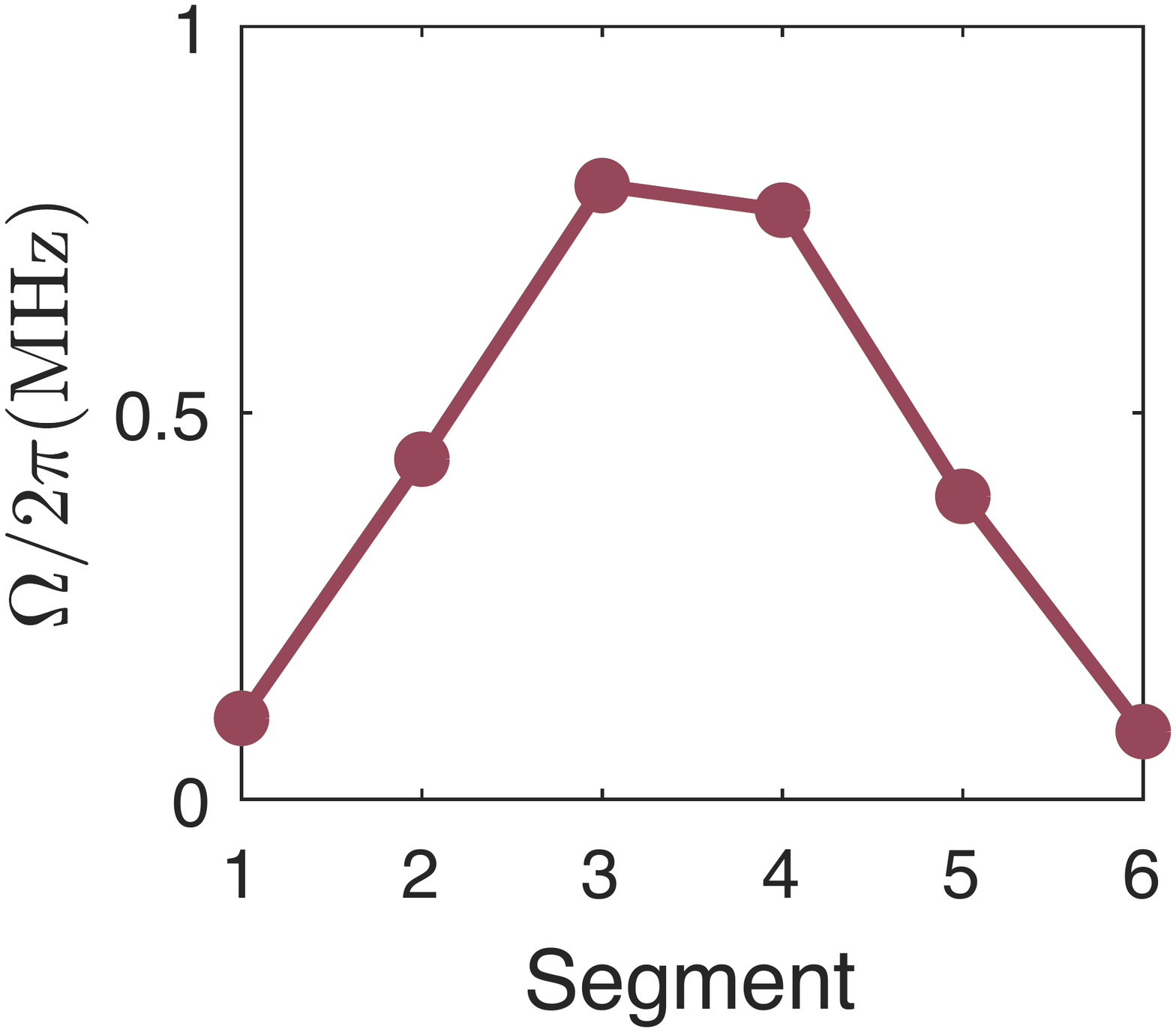}}
\caption{(a) Optimal gate infidelity vs. laser detuning $\mu$
for $n=6$ segments and a total gate time $\tau=50\,\mu$s. (b) Rabi frequencies
for each segment when $\mu$ is chosen at the minimizer of (a).
The gate design is optimized through amplitude modulation \cite{wu_noise_2018}. There are 3
curves in each plot, blue for $N=50$ ions and ion pair 25 and 26, green for
$N=50$ ions and ion pair 10 and 11, red for $N=100$ ions and ion pair
50 and 51. All these curves coincide within the resolution of the figure.
Here we choose an ion spacing $d=8\,\mu$m, Lamb-Dicke parameter $g_0=0.1$,
transverse trapping frequency $\omega_1=2\pi\times 3\,$MHz and 0.5 phonon per mode.}\label{fig:gate_design}
\end{figure}

In this section we present some numerical results for a 1D ion chain. We have proven that the distant ions have little effect on the fidelity of an entangling gate, which suggests that when designing the gate we can ignore all the ions far away and focus on a finite number of ions. This is verified numerically by the example shown in Fig.~\ref{fig:gate_design}, where we consider gate design in uniform ion chains with $d=8\,\mu$m but varying total number of ions and the position of the ion pair inside the chain. The optimal gate design and the intrinsic infidelity for these cases are almost identical as they overlap with each other in the plot.

\begin{figure}[htbp]
\centering
\subfigure[]{\includegraphics[width=40mm]{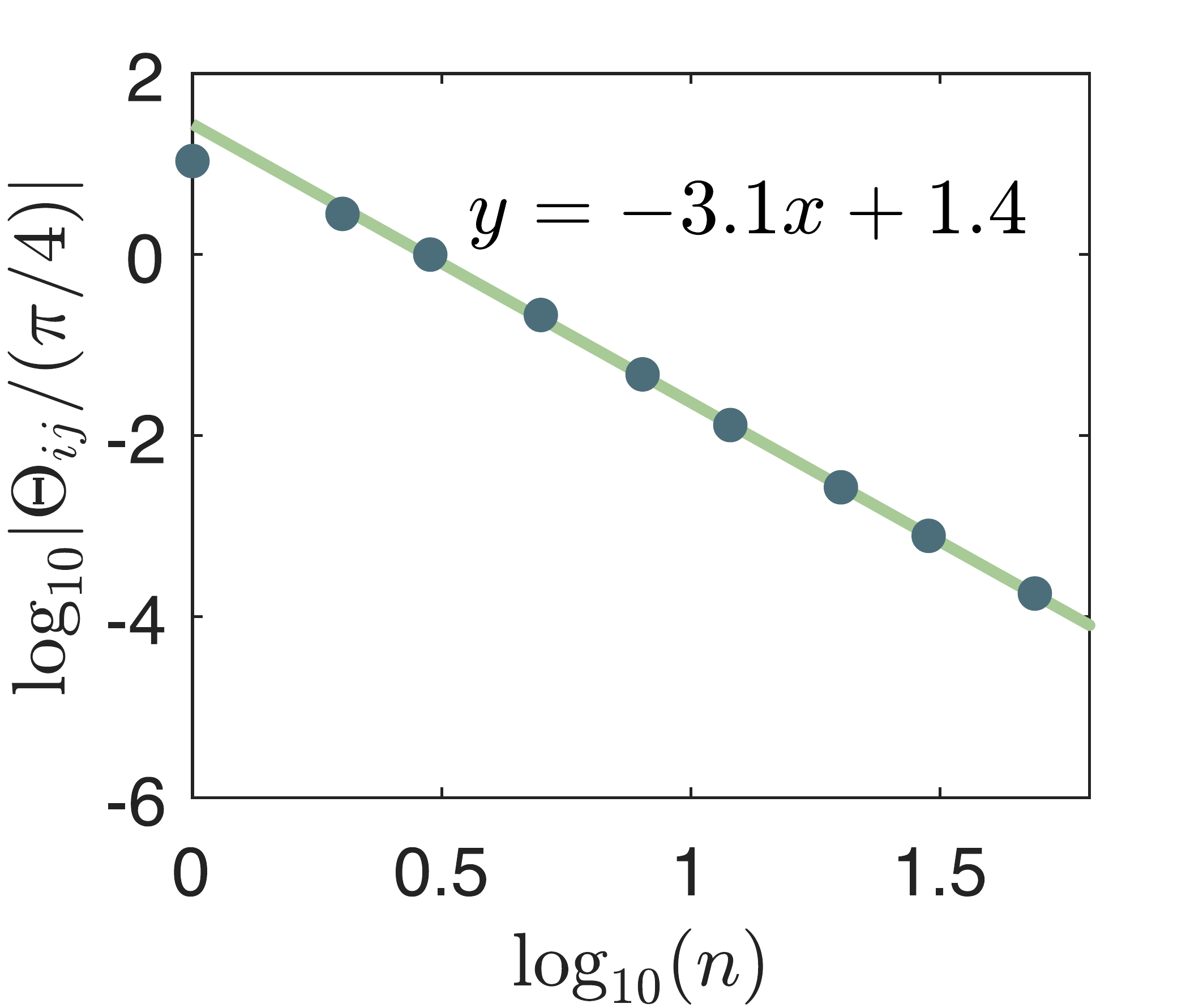}}
\subfigure[]{\includegraphics[width=40mm]{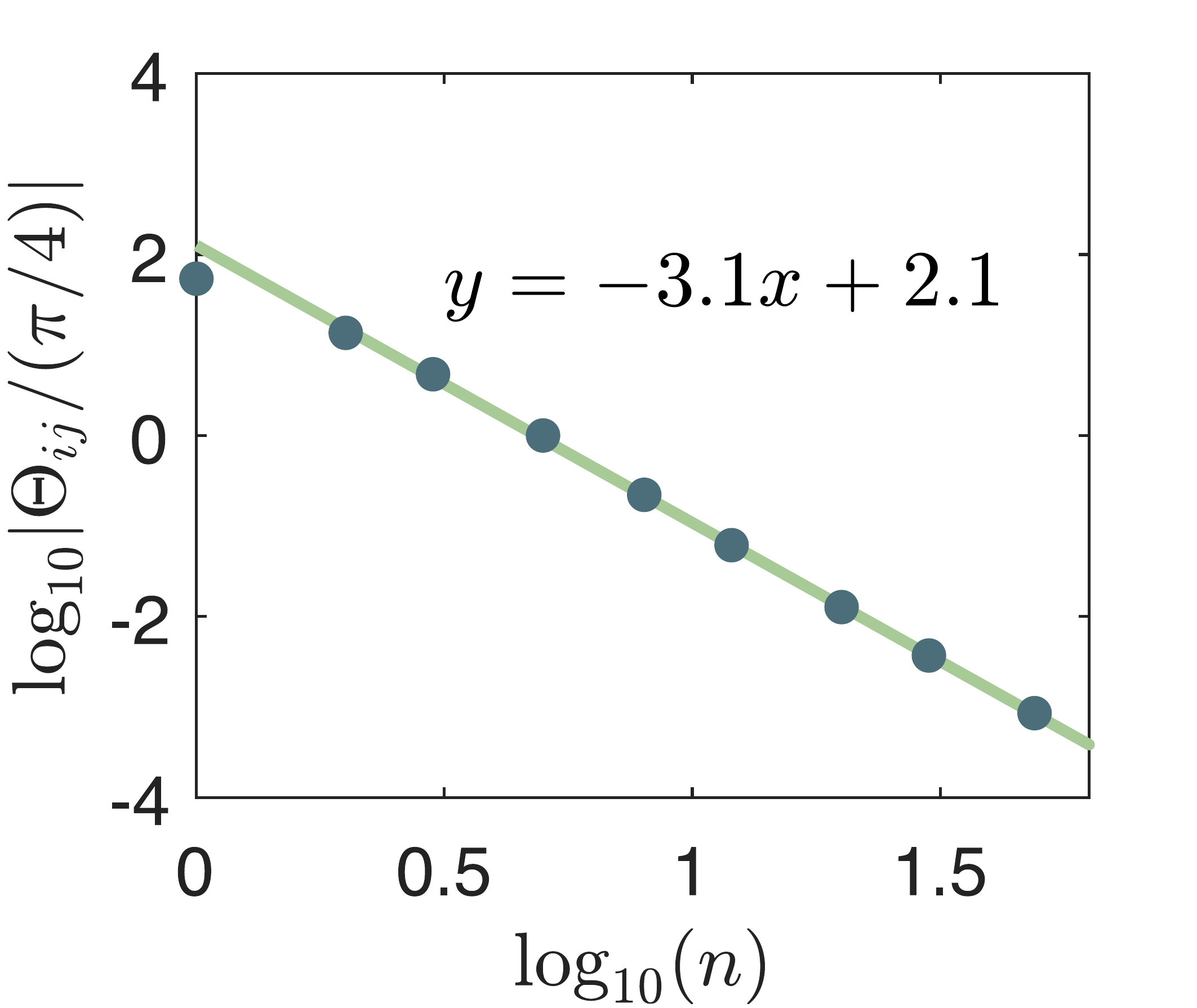}}
\caption{$|\Theta_{i,j}|$ vs. $n=|j-i|$ for
(a) a gate designed for two ions at a spacing of 3 separations with $\tau=60\,\mu$s, $n_{\mathrm{seg}}=7$ segments, $\mu=1.01403\omega_1$,
(b) a gate designed for two ions at a spacing of 5 separations with $\tau=100\,\mu$s, $n_{\mathrm{seg}}=10$ segments, $\mu=1.01387\omega_1$.
The green lines are fitted from the last five data points in each figure.}\label{fig:scaling_m35}
\end{figure}

In the main text we show the scaling of the crosstalk error, $\|\mathcal{E}\|_\diamond$, vs. gate distance, $n$, for a gate designed for a nearest-neighbor ion pair. Here we plot results for entangling gates designed for ion pairs at a distance of 3 and 5 in Fig.~\ref{fig:scaling_m35} and observe a similar scaling of $1/n^3$. For simplicity we directly plot $\Theta_{i,j}$ vs. $n=|j-i|$ while $\|\mathcal{E}\|_\diamond$ can be bounded by four such $(i,\,j)$ pairs (see the main text).

\end{document}